\documentclass[preprint,12pt,number]{elsarticle}

\usepackage{graphicx}
\usepackage{amssymb}
\usepackage{amsthm}
\usepackage{amsmath}

\journal{Nuclear Physics A}

\begin{document}

\begin{frontmatter}

\title{Fission of heavy 
$\Lambda$ hypernuclei with the Skyrme-Hartree-Fock approach}

\author[JAEA]{F. Minato}
\ead{minato.futoshi@jaea.go.jp}
\author[JAEA,NAO]{S. Chiba}
\author[TOHOKU]{K. Hagino}

\address[JAEA]{Japan Atomic Energy Agency, Tokai 319-1195, Japan}
\address[NAO]{National Astronomical Observatory of Japan, Osawa, Mitaka,
Tokyo 181-8588, Japan}
\address[TOHOKU]
{Department of Physics, Tohoku University, Sendai 980-8578, Japan}

\begin{abstract}
Fission-related phenomena
of heavy $\Lambda$ hypernuclei are discussed 
with the constraint Skyrme-Hartree-Fock+BCS (SHF+BCS) method, in 
which a similar Skyrme-type interaction is employed also 
for the interaction between a $\Lambda$ particle and a nucleon. 
Assuming that the $\Lambda$ particle 
adiabatically follows the fission motion, 
we discuss the fission barrier height of $^{239}_{\;\;\;\Lambda}$U. 
We find that the fission barrier height increases slightly 
when the $\Lambda$ particle occupies the lowest level. 
In this case, 
the $\Lambda$ particle is always attached to the heavier fission fragment. 
This indicates that 
one may produce heavy neutron-rich $\Lambda$ hypernuclei 
through fission, whose weak decay is helpful for 
the nuclear transmutation of long-lived fission products.
We also discuss cases where the $\Lambda$ particle occupies a higher 
single-particle level. 
\end{abstract}


\end{frontmatter}

\section{Introduction}
\label{introduction}

Hyperons have attracted much interests since its discovery.
The interaction between a hyperon ($Y$) and a nucleon ($N$) is
important not only for 
understanding the mechanism of nuclear force but also for 
the equation of state of nuclear matter in astrophysical sites. 
$YN$ scattering experiments have been difficult to perform 
since 
the life-time of hyperons in vacuum is so short (typically 
a few $10^{-10}$ seconds). 
Therefore,
the information on the $YN$ interaction has been obtained 
mainly from the structure of finite hypernuclei, which are 
composed of neutrons, protons and at least one hyperon. 
There have been studied by now 
many single-$\Lambda$ hypernuclei from H to U \cite{HT06}, 
and a few $\Lambda\Lambda$ and $\Sigma$ hypernuclei have 
also been produced experimentally.

One of the important questions in hypernuclear physics 
is to clarify the role of hyperon 
in the structure of hypernuclei. 
An attractive $YN$ force 
may cause a significant modification in the properties of a finite nucleus 
when a hyperon is added to it. 
Such impurity effect of hyperon for ${}^{7}_\Lambda$Li, that is, 
the shrinkage of ${}^6$Li core in the hypernucleus, 
was predicted 
theoretically using the cluster models \cite{Bando1983,Hiyama1996,Hiyama1999}, 
which has subsequently been 
confirmed experimentally by Tanida {\it et al.} \cite{Tanida2001}.
A recent theoretical work 
with the relativistic mean field (RMF) theory 
has also shown that the shape of 
$^{12}$C and $^{28}$Si changes from oblate to spherical 
when a $\Lambda$ particle is added to them \cite{Myaing2009}.

It is intriguing to ask whether the similar effects 
can be expected for heavy fissile nuclei.
Interests in the fission of heavy $\Lambda$ hypernuclei include 
i) a change of fission barrier height due to the additional $\Lambda$ 
particle and ii) 
an attachment probability of 
a $\Lambda$ particle to each fission fragments after fission takes place. 
The latter has been 
studied experimentally by Armstrong {\it et al.} 
for hypernuclei formed in anti-proton annihilation 
on $^{209}$Bi and $^{238}$U nuclei 
\cite{Armstrong1993}.
They reported that the $\Lambda$ particle predominantly 
sticks to the heavier fission fragment.
Theoretical works on the $\Lambda$ attachment probability 
have been also performed in Refs. \cite{Krappe1993, Karpeshin1995, Krappe1996}
with a phenomenological Woods-Saxon potential and the statistical method,
yielding consistent results with 
the experiment finding. 
As for the former interest, to our knowledge, 
the fission barrier height of heavy hypernuclei 
has been studied neither theoretically nor experimentally.

The aim of this paper is then to clarify the impurity effects of $\Lambda$ 
particle on
the fission barrier height of heavy $\Lambda$ hypernuclei and 
investigate a possibility to 
produce 
neutron-rich $\Lambda$ hypernuclei, 
which are difficult to produce directly by experimental beams.
To this end, 
we use the (constraint) Skyrme-Hartree-Fock+BCS (SHF+BCS) method.
This method has been widely used for the study of fission 
\cite{SkM,Nazarewicz,Minato2008,Bender}
and has been applied to hypernuclei as well 
\cite{Rayet,Yamamoto1988,Yamamoto1990,Lanskoy1997,
Lanskoy1998,Cugnon2000,Vidana2001}.
It is also a virtue of this method that 
one can calculate the wave functions of $\Lambda$ particle
in a self-consistent manner. 

The paper is organized as follows.
In section 2, we describe the theoretical framework of 
the constraint SHF+BCS method.
In section 3, we illustrate the potential 
energy surface for fission of heavy hypernuclei and
discuss an evolution of $\Lambda$ particle motion during fission. 
In section 4, 
we give the conclusion and discuss possible applications 
of fission of heavy $\Lambda$ hypernuclei.

\section{Skyrme-Hartree-Fock method for hypernuclei}
\label{Theory}

The Skyrme-Hartree-Fock method \cite{Vautherin1972} 
has been extensively used for stable and unstable nuclei.
In this approach, the effective $NN$ interaction is described by the Skyrme interaction,
which has the form of density-dependent zero-range interaction.

This method has been extended to hypernuclei \cite{Rayet}, 
in which the $\Lambda N$ and $\Lambda NN$ interactions are also described by
the similar Skyrme-like $\delta$-interaction.  
The two-body $\Lambda N$ interaction is given by,
\begin{equation}
\begin{split}
v_{\Lambda N}(\vec{r}_\Lambda-\vec{r}_N)
&=
t_0^\Lambda(1+x_0^\Lambda P_\sigma)\delta(\vec{r}_\Lambda-\vec{r}_N)\\
&
+\frac{1}{2}t_1^\Lambda
\left(
\vec{k'}^2 \delta( \vec{r}_\Lambda - \vec{r}_N )
+\delta( \vec{r}_\Lambda - \vec{r}_N )\vec{k}^2 \right)\\
&
+t_2^\Lambda \vec{k'} \delta(\vec{r}_\Lambda-\vec{r}_N)\cdot \vec{k}
+iW_0^\Lambda\vec{k'}\delta(\vec{r}_\Lambda-\vec{r}_N)\cdot(\sigma\times\vec{k}),
\end{split}
\label{2body}
\end{equation}
while the three-body $\Lambda NN$ interaction is,
\begin{equation}
v_{\Lambda NN}(\vec{r}_\Lambda,\vec{r}_1,\vec{r}_2)
=t_3^\Lambda\delta(\vec{r}_\Lambda-\vec{r}_1)\delta(\vec{r}_\Lambda-\vec{r}_2).
\label{3body}
\end{equation}
Here, $\vec{k}$ and $\vec{k}'$ are the derivative operators acting 
on the right and the left hand sides, respectively. 
The parameters in Eqs. \eqref{2body} and \eqref{3body} have been 
adjusted to fit 
the experimental binding energies of $\Lambda$ hypernuclei 
\cite{Rayet,Yamamoto1988}. 
With these effective interactions, the total energy $E$ of hypernucleus 
is given by,
\begin{equation}
E=E_N+E_\Lambda
\end{equation}
\begin{equation}
\begin{split}
E_N       &=\int  H_N(\vec{r}) \, d\vec{r} + E_N^{\rm{pair}}\\
E_\Lambda &=\int  H_\Lambda(\vec{r}) \, d\vec{r},
\end{split}
\label{total_energy}
\end{equation}
where $E_N$, $E_\Lambda$, and $E_N^{\rm{pair}}$ are 
the energy of the core nucleus, that of the $\Lambda$ particle, 
and the pairing energy, respectively.
The energy density for the core nucleus, $H_N(\vec{r})$, 
is standard and its explicit form can be found {\it e.g.,} 
in Refs.\cite{Vautherin1972,Minato2008}. 
The energy density for the $\Lambda$ particle, $H_\Lambda(\vec{r})$, is 
given by, 
\begin{equation}
\begin{split}
H_\Lambda(\vec{r})
&
=\frac{\hbar^2}{2m_\Lambda}\tau_\Lambda
+t_0^\Lambda \Big(1+\frac{1}{2}x_0^\Lambda \Big) \rho\,\rho_\Lambda
+\frac{1}{4}\Big(t_1^\Lambda+t_2^\Lambda \Big)
\Big(\tau_\Lambda\,\rho\,+\tau\,\rho_\Lambda\Big)\\
&
+\frac{1}{4}\Big(3t_1^\Lambda-t_2^\Lambda\Big)
\Big(\vec{\nabla}\rho\,\cdot\vec{\nabla}\rho_\Lambda\Big)
+\frac{1}{2}W_0^\Lambda 
\Big(\vec{\nabla}\rho\cdot\vec{J}_\Lambda+\vec{\nabla}\cdot\vec{J}\Big)\\
&
+\frac{1}{4}t_3^\Lambda\rho_\Lambda
\Big(\rho^2+2\rho_n\,\rho_p \Big),
\end{split}
\end{equation}
where $\rho=\rho(\vec{r})=\rho_p(\vec{r})+\rho_n(\vec{r})$, 
$\tau=\tau(\vec{r})=\tau_p(\vec{r})+\tau_n(\vec{r})$, 
and 
$\vec{J}=\vec{J}(\vec{r})=\vec{J}_p(\vec{r})+\vec{J}_n(\vec{r})$ 
are the number, the kinetic energy, and the spin-current 
densities for the core nucleus, respectively. 
$\rho_\Lambda=\rho_\Lambda(\vec{r})$, 
$\tau_\Lambda=\tau_\Lambda(\vec{r})$, 
and $\vec{J}_\Lambda=\vec{J}_\Lambda(\vec{r})$ 
are the same quantities, but for the $\Lambda$ particle. 

The single-particle wave functions $\phi(\vec{r})$ are obtained by minimizing 
the total energy $E$. This leads to the 
SHF equations for proton ($q=p$) and neutron ($q=n$) given by 
\begin{equation}
\left(
-\vec{\nabla}\cdot \frac{\hbar^2}{2m_q^*(\vec{r})}
\vec{\nabla}+U_q(\vec{r})+U_q^\Lambda(\vec{r})
\right)
\phi_q(\vec{r}) = \epsilon_q \phi_q(\vec{r}), 
\end{equation}
where $m_q^*(\vec{r})$ is the nucleon effective mass defined as,
\begin{equation}
\frac{\hbar^2}{2m_q^*(\vec{r})}
=\frac{\hbar^2}{2m_q}
+\frac{1}{4}\Big(t_1+t_2\Big)\rho(\vec{r})
+\frac{1}{8}\Big(t_2-t_1\Big)\rho_q(\vec{r})
+\frac{1}{4}(t_1^\Lambda+t_2^\Lambda) \rho_\Lambda(\vec{r}). 
\end{equation}
$U_q^\Lambda(\vec{r})$ is the mean-field potential
due to the $\Lambda N$ interaction 
given by,
\begin{equation}
\begin{split}
U_q^\Lambda(\vec{r}) 
&= t_0^\Lambda\left(1+\frac{1}{2}x_0^\Lambda\right) 
\rho_\Lambda(\vec{r}) + \frac{1}{4}(t_1^\Lambda+t_2^\Lambda) \tau_\Lambda(\vec{r})
 -\frac{1}{4}(3t_1^\Lambda-t_2^\Lambda) \vec{\nabla}^2\rho_\Lambda(\vec{r})\\
&-\frac{1}{2}W_0^\Lambda\left(\vec{\nabla}\vec{J}_\Lambda(\vec{r})\right)
 +\frac{1}{2}W_0^\Lambda\vec{\nabla}\rho_\Lambda(\vec{r})(-i)
 (\vec{\nabla}\times\vec{\sigma})
 +\frac{1}{2}t_3^\Lambda\rho_\Lambda(\vec{r})\Big(\rho(\vec{r})+\rho_q(\vec{r})\Big).
\end{split}
\end{equation}
Similarly, the wave function for the $\Lambda$ particle 
is obtained with the equation, 
\begin{equation}
\left(
-\vec{\nabla}\cdot \frac{\hbar^2}{2m_\Lambda^*(\vec{r})}
\vec{\nabla}+U_\Lambda^{N}(\vec{r})
\right)
\phi_\Lambda(\vec{r})=\epsilon_\Lambda\phi_\Lambda(\vec{r}),
\end{equation}
where $m_\Lambda^*(\vec{r})$ are the $\Lambda$ effective mass defined as,
\begin{equation}
\frac{\hbar^2}{2m_\Lambda^*(\vec{r})}
=\frac{\hbar^2}{2m_\Lambda}
+\frac{1}{4}\Big(t_1^\Lambda+t_2^\Lambda\Big)\rho(\vec{r}).
\end{equation}
The mean-field potential $U_\Lambda^{N}$ for 
the $\Lambda$ particle is given by,
\begin{equation}
\begin{split}
U_\Lambda^{N}(\vec{r})
&= t_0^\Lambda\left(1+\frac{1}{2}x_0^\Lambda\right)\rho(\vec{r})
  +\frac{1}{4}(t_1^\Lambda+t_2^\Lambda) \tau(\vec{r})
  -\frac{1}{4}(3t_1^\Lambda-t_2^\Lambda) \vec{\nabla}^2\rho(\vec{r})\\
& +\frac{1}{2}W_0^\Lambda\vec{\nabla}\rho(\vec{r})(-i)
  (\vec{\nabla}\times\vec{\sigma})
 -\frac{1}{2}W_0^\Lambda\vec{\nabla}\vec{J}(\vec{r})
 +\frac{1}{4}t_3^\Lambda\Big(\rho^2(\vec{r})+2\rho_n(\vec{r})\rho_p(\vec{r})\Big).
\end{split}
\end{equation}

We define the quadrupole operator as,
\begin{equation}
\hat{Q}^{q}_{2}
=\sqrt{\frac{16\pi}{5}}\sum_{i\in q}
r_i^2Y_{20}(\theta_i),
\end{equation}
and the octupole operator as,
\begin{equation}
\hat{Q}^{q}_{3}
=\sum_{i\in q}
r_i^3Y_{30}(\theta_i),
\end{equation}
respectively ($q=p,n,\Lambda$).
Using the self-consistent solution of the SHF equations, 
the multipole moments are calculated as,
\begin{eqnarray}
Q^{q}_2&=&
\sqrt{\frac{16\pi}{5}}\,
\int d\vec{r}\,
r^2Y_{20}(\theta)\rho_q(\vec{r}), \\
Q^{q}_3&=&
\int d\vec{r}\,
r^3Y_{30}(\theta)\rho_q(\vec{r}),
\end{eqnarray}
for the quadrupole and octupole, respectively. 

We perform our calculations 
by modifying the 
computer code {\tt SKYAX} \cite{SKYAX,Reinhard1999}. 
This code solves the SHF equations in the coordinate space 
assuming axial symmetry for the nuclear shape.
The potential energy surface for fission 
is obtained by constraining on the quadrupole moment, while 
the octupole moment is optimized 
by adding a small octupole moment to the initial wave functions. 
We take the mesh size of $\Delta r=\Delta z=1.0$ fm 
with 
boundary conditions of vanishing wave functions at 
$r=20$ fm, $z=-25$ fm, and $z$=25 fm, where $r=\sqrt{x^2+y^2}$. 

The pairing correlation among nucleons 
is treated in the BCS approximation.
We use the density-dependent contact interaction,
\begin{equation}
v_{\rm pair}^q(\vec{r}_1,\vec{r}_2) 
= v_0^q\left(1-\frac{\rho(\vec{r})}{\rho_0}\right)\delta(\vec{r}_1-\vec{r}_2),
\end{equation}
for the pairing interaction. 
In the code, a smooth cut-off function \cite{Bender1998}
\begin{equation}
f_k = \frac{1}{
1 + \exp((\epsilon_k-\lambda-\Delta E)/\mu)}
\end{equation}
is introduced for the pairing active space. Here, $\lambda$ is the
Fermi energy, and $\Delta E$ is determined so that
\begin{equation}
N_{\rm{act}}=\sum_k
f_k = N_q + 1.65 N^{2/3}_q
\end{equation}
with $\mu=\Delta E/10$, $N_q$ being the number of proton or neutron.
The pairing energy is given by,
\begin{equation}
E_N^{\rm{pair}}= \sum_{k\in p,n} u_k v_k \,\Delta_k f_k,
\end{equation}
where $u_k$ and $v_k$ are the $uv$ factor in the BCS approximation, 
and $\Delta_k$ is the pairing gap.

\section{Results}

Heavy $\Lambda$ hypernuclei, which are produced by 
($K^-$,$\pi^-$), ($\pi^+$,$K^+$), or ($e$,$e'K^+$) reactions,
are initially at an excited state. 
Then the fission channel competes with several other decay channels.
Three fission processes can be considered: 
(I) the $\Lambda$ particle itself decays via 
non-mesonic decay followed by fission of the remaining nucleus, 
(I\hspace{-.1em}I) 
fission occurs after the $\Lambda$ particle
de-excites to the lowest single-particle state 
emitting several $\gamma$-rays, 
(I\hspace{-.1em}I\hspace{-.1em}I) 
fission occurs while the $\Lambda$ particle is at an excited single-particle 
level.
Among them, the process (I) corresponds to a statistical fission of an 
excited nucleus and we do not consider it in this work.

We calculate the potential 
energy curve of $^{239}_{\;\;\;\Lambda}$U as a function of 
$Q_2=Q_2^p+Q_2^n +Q_2^\Lambda$,
and compare it with that of $^{238}$U.
We set the pairing strength
parameters to be $v_0^p = 1410.0$ MeV$\cdot$fm$^3$ for proton and
$v_0^n=910.0$ MeV$\cdot$fm$^3$ for neutron so as to reproduce
the empirical pairing gap of $^{238}$U, that is,  
$\Delta_n=0.674$ MeV and $\Delta_p=1.168$ MeV obtained with the 
three-point formula for nuclear mass.
We adopt the parameter set SkM$^*$ \cite{SkM} for the $NN$ interaction, 
while the parameter No. 4 in Table. I of Ref. \cite{Yamamoto1988} 
for the $\Lambda N$ interaction, 
which is optimized to $^{209}_{\;\;\;\Lambda}$Pb,
\begin{center}
\begin{tabular}{ccc}
  $t_0^\Lambda$  $=-253.500$ MeV$\cdot$fm$^3$,
& $x_0^\Lambda$  $=  -0.216$,
& $t_1^\Lambda$  $=  76.920$ MeV$\cdot$fm$^5$,\\
  $t_2^\Lambda$  $=  28.080$ MeV$\cdot$fm$^5$,
& $t_3^\Lambda$  $=   0.000$ MeV$\cdot$fm$^6$,
& $W_0^\Lambda$  $=   0.000$ MeV$\cdot$fm$^5$. 
\end{tabular}
\end{center}
We have confirmed that 
the potential energy curve of $^{239}_{\;\;\;\Lambda}$U
does not alter significantly 
even when other parameter sets for the $\Lambda N$ interaction 
\cite{Rayet, Yamamoto1988} are employed.

\subsection{$\Lambda$ at the lowest single-particle level}

\begin{figure}
\begin{center}
\includegraphics[width=0.60\textwidth]{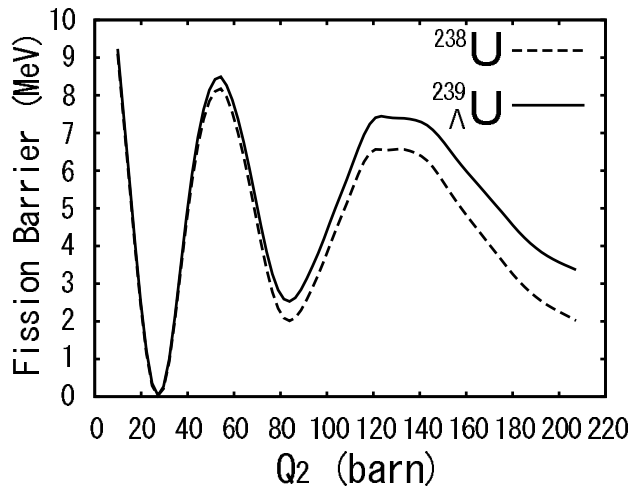}
\caption{The fission barrier of $^{238}$U (the dotted line) 
and $^{239}_{\;\;\;\Lambda}$U (the solid line) nuclei obtained with 
the Skyrme-Hartree-Fock method. The $\Lambda$ particle 
is assumed to occupy the lowest single-particle state during 
fission. The energy curves are shifted so that the ground state 
configuration has zero energy. }
\label{bar1}
\end{center}
\end{figure}

Let us first discuss the fission of hypernucleus $^{239}_{\;\;\;\Lambda}$U 
at the ground state. 
For this purpose, we assume 
that a $\Lambda$ particle follows adiabatically the fission motion.
That is, the $\Lambda$ particle 
is at the lowest single-particle level at every instant.
Fig. \ref{bar1} shows the fission barrier for 
$^{238}$U (the dashed line) and $^{239}_{\;\;\;\Lambda}$U (the solid line)
as a function of total quadrupole moment $Q_2$.
Those curves are shifted so that the ground state configuration 
has zero energy. 
The deformation parameter for the ground state 
does not change significantly by adding a 
$\Lambda$ particle. 
On the other hand, 
the fission barrier height of $^{239}_{\;\;\;\Lambda}$U 
slightly increases as compared to that of $^{238}$U.
This suggests that $^{238}$U 
may become more stable by adding a $\Lambda$ particle, although 
one needs to evaluate also the effect of $\Lambda$ particle on 
the mass inertia to reach a definite conclusion. 
We list the height of the inner and outer fission barriers in Tab. \ref{bar1}.
The inner barrier height increases by about $0.27$ MeV, while  
the outer barrier by about $0.82$ MeV due to the additional 
$\Lambda$ particle in the ${}^{239}_{\;\;\;\Lambda}$U 
nucleus. 

\begin{table}
\begin{center}
\begin{tabular}{c|cc}
\hline
     & \quad $^{238}$U \quad & $^{239}_{\;\;\;\Lambda}$U\\
\hline
$B_f$(inner) (MeV) & $8.20$    & $8.47$ \\
$B_f$(outer) (MeV) & $6.60$    & $7.42$ \\
\hline
\end{tabular}
\label{Bf1}
\caption{
The height of the inner and outer fission barriers for the 
$^{238}$U and $^{239}_{\;\;\;\Lambda}$U nuclei when 
the $\Lambda$ particle occupies the lowest single-particle state during 
fission. 
}
\end{center}
\end{table}

\begin{figure}
\begin{center}
\includegraphics[width=0.50\textwidth]{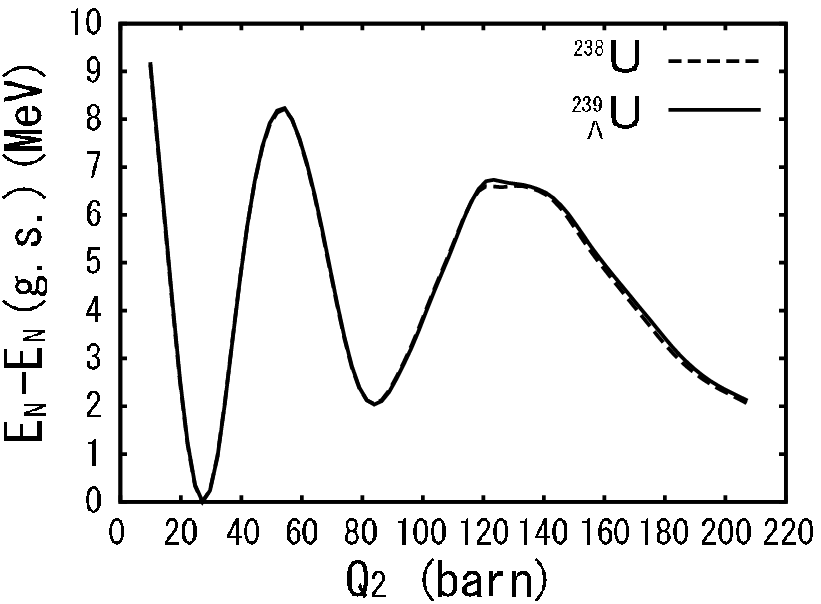}
\includegraphics[width=0.50\textwidth]{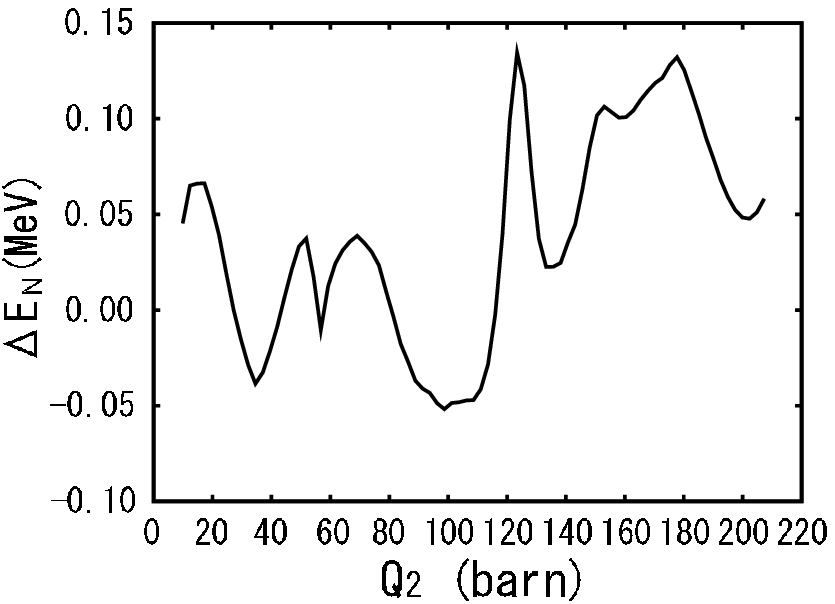}
\includegraphics[width=0.50\textwidth]{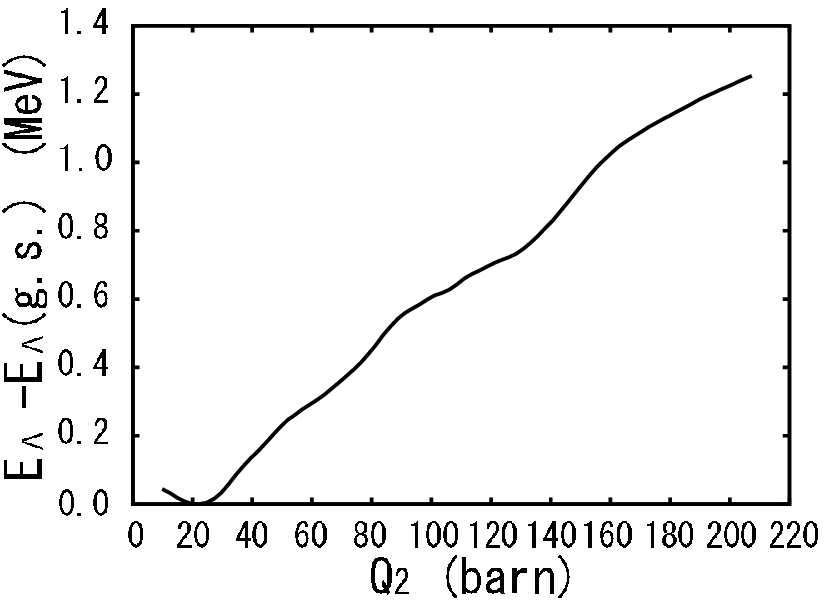}
\caption{
(The top panel:)
The energy of the core nucleus $E_N$ 
for 
the $^{238}$U (the solid line) and $^{239}_{~~\Lambda} $U (the dashed line) 
nuclei as a function of the total quadrupole moment $Q_2$. 
The $\Lambda$ particle is assumed to 
be at the lowest single-particle state. 
(The middle panel:)
The difference between the solid and the dashed curves in the 
upper left panel. 
(The bottom panel:)
The energy of the $\Lambda$ particle $E_\Lambda$ 
for $^{239}_{~~\Lambda}$U 
with respect to that 
for the ground state as a function of $Q_2$. 
}
\label{core}
\end{center}
\end{figure}

The degree of increase of the fission barrier height
is primarily determined by the energy of $\Lambda$ particle, $E_\Lambda$.
The fission barrier height is defined as 
the energy difference at
the saddle point configuration (s.p.) and the ground state (g.s.). 
This reads
\begin{equation}
\begin{split}
B_f
&=E({\rm s.p.})-E({\rm g.s.})\\
&=\Big( E_N({\rm s.p.})-E_N({\rm g.s.}) \Big) 
+ \Big( E_\Lambda({\rm s.p.})-E_\Lambda({\rm g.s.}) \Big).
\end{split}
\label{fissionbarrier}
\end{equation}
The first and second terms in Eq. \eqref{fissionbarrier} correspond to
the energy difference for the core nucleus 
and the $\Lambda$ particle, respectively.
We plot these quantities as a function of $Q_2$ in Fig. \ref{core}.
The top panel shows $E_N-E_N(\rm{g.s.})$ 
for $^{238}$U (the solid line) and $^{239}_{~~\Lambda} $U (the dashed line). 
Notice that 
the difference between these two curves is small, 
as shown in the middle panel in Fig. \ref{core}. 
The difference is in fact about the order of 1\% of the barrier height, 
and thus we conclude that 
the energy of the core nucleus is insensitive 
to the presence of $\Lambda$ particle. 
In contrast, the energy of the $\Lambda$ particle, $E_\Lambda$, varies 
more significantly as a function of $Q_2$. 
The bottom panel in Fig. \ref{core} shows 
$E_\Lambda-E_\Lambda(\rm{g.s.})$ as a function of $Q_2$.
The energy $E_\Lambda$ monotonically increases with the quadrupole moment $Q_2$
with respect to the ground state.
The energy difference 
is $\Delta E_\Lambda=E_\Lambda({\rm s.p.})-E_\Lambda({\rm g.s.})=0.25$ MeV 
at the inner barrier position, which 
almost amounts to the increase of the fission 
barrier height ($\Delta B_f=0.27$ MeV).
It is thus evident that 
the change in the fission barrier height is dominantly 
resulted from the energy of the $\Lambda$ 
particle. 

\begin{figure}
\begin{center}
\includegraphics[height=1.00\textwidth]{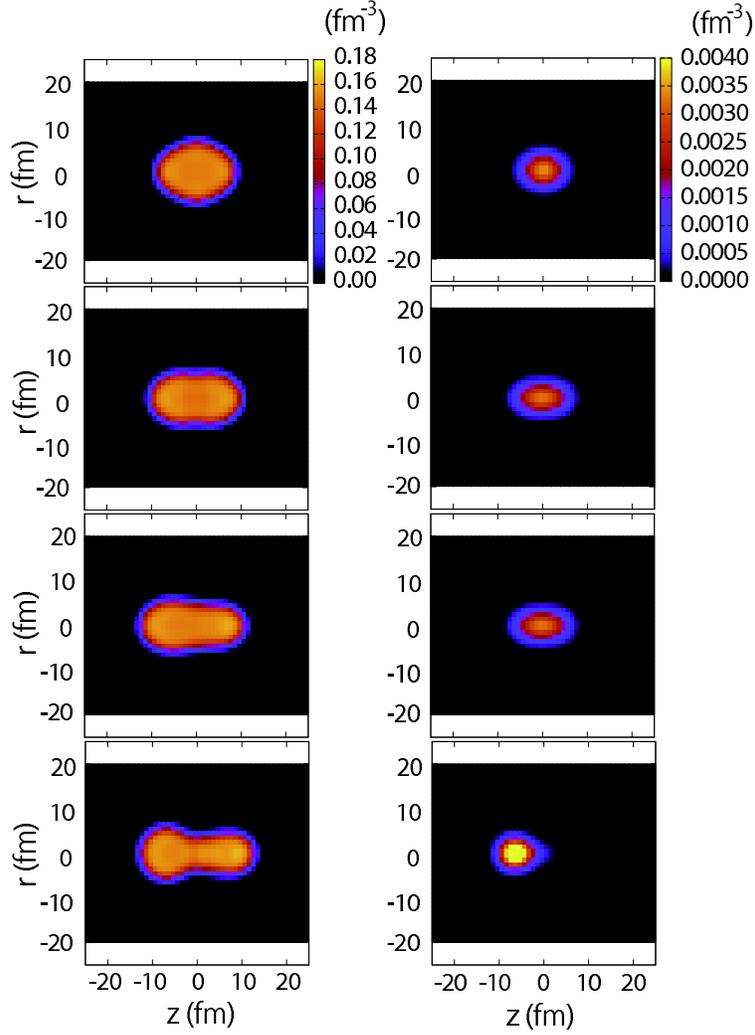}
\caption{
The density distribution for the core nucleus (the left panels) 
and the $\Lambda$ particle (the right panels) in 
$^{239}_{~~\Lambda} $U 
at the ground state ($Q_2=27.15$ b), the second minimum ($Q_2=83.90$ b), 
the outer saddle point ($Q_2=123.39$ b), and $Q_2=200$ b.
The $\Lambda$ particle is put at the lowest single particle level
at all the deformations.}
\label{density1}
\end{center}
\end{figure}

Fig. \ref{density1} shows 
the density distributions for the core nucleus $^{238}$U (the left panels)
in $^{239}_{~~\Lambda}$U 
and the $\Lambda$ particle (the right panels)
at the ground state ($Q_2=27.15$ b), the second minimum ($Q_2=83.90$ b), 
the outer saddle point ($Q_2=123.39$ b), and $Q_2=200.00$ b.
At $Q_2=200$ b,
the core nucleus is separated asymmetrically into two nuclei, 
where 
the fragment on the left-hand side is heavier than that on the 
right-hand side (the initial octupole moment at the first stage of 
iteration determines which fragment is heavier).
We see that the $\Lambda$ particle is 
localized in the region of $z<0$, that is, it is 
stuck to the heavier fission fragment. 
This is a natural consequence of the fact 
that the binding energy for the $\Lambda$ particle 
is larger in the heavier nucleus. 

\begin{figure}
\begin{center}
\includegraphics[width=0.55\textwidth]{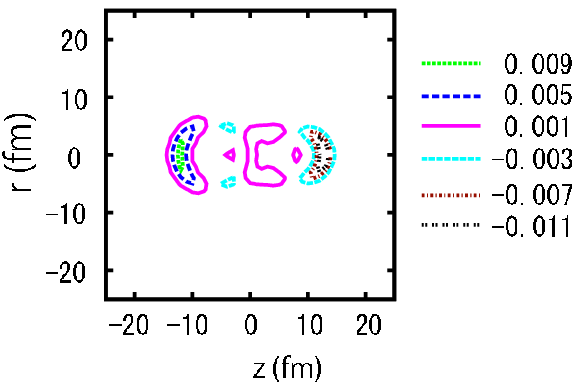}
\caption{
The difference between the density 
of the core nucleus $^{238}$U 
in $^{239}_{~~\Lambda} $U and the density 
of $^{238}$U in the absence of the $\Lambda$ particle 
at $Q_2=200$ barn. }
\label{mdist1}
\end{center}
\end{figure}

Notice that the $\Lambda N$ attractive interaction can 
attract more nucleons in the heavier fission fragment. 
In order to see this effect,
we plot in Fig. \ref{mdist1} 
the difference between the density 
of the core nucleus $^{238}$U 
in $^{239}_{~~\Lambda} $U and the density 
of $^{238}$U in the absence of the $\Lambda$ particle 
at $Q_2=200$ barn. 
That is, 
\begin{equation}
\Delta \rho(\vec{r})=\rho^\Lambda_{\rm{core}}(\vec{r})-\rho(\vec{r}),
\end{equation}
where $\rho_{\rm{core}}^\Lambda(\vec{r})$ is 
the density of $^{238}$U in ${}^{239}_{\;\;\;\Lambda}$U 
and $\rho(\vec{r})$ is the density of $^{238}$U without the 
$\Lambda$ particle.
As shown in Fig. \ref{density1}, 
the left hand side ($z<0$) corresponds to the heavier fragment, 
to which the $\Lambda$ particle is stuck.
We can see that $\Delta \rho(\vec{r})$ is positive (red) on the left side,
and nucleons are actually attracted into the heavier fragment 
by the presence of $\Lambda$ particle.
If we integrate the region of $z<0$, about 0.54 proton and 0.86 neutrons 
are attracted due to the $\Lambda$ particle.

\begin{figure}
\begin{center}
\includegraphics[width=0.49\textwidth]{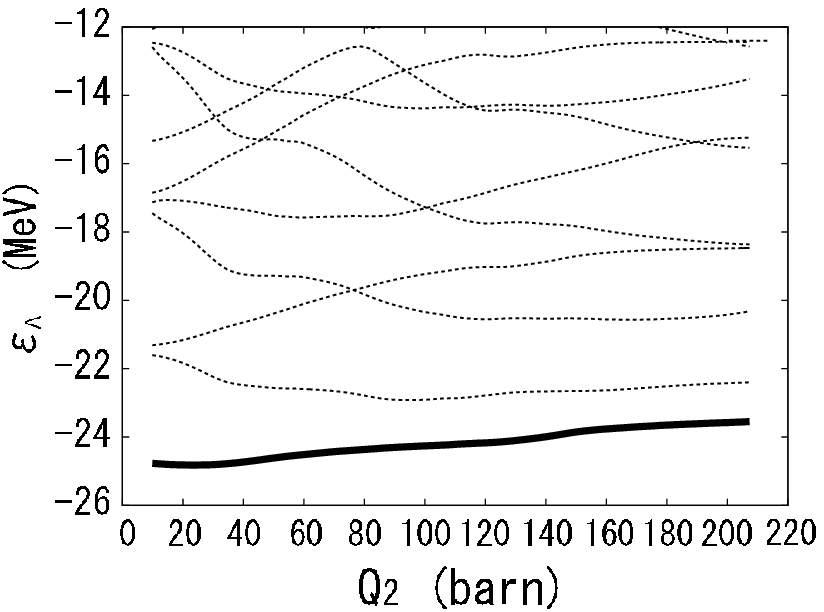}
\caption{Single-particle levels for the $\Lambda$ particle 
in $^{239}_{~~\Lambda}$U as a function of $Q_2$.
We assume that the $\Lambda$ particle stays at the lowest level 
(the thick solid line)
at all the deformations.}
\label{sp1}
\end{center}
\end{figure}

The motion of the $\Lambda$ particle during fission 
can also be inferred from 
the single-particle levels for the $\Lambda$ particle. 
We plot them in Fig. \ref{sp1} as a function of $Q_2$.
The lowest level (the thick solid line) 
does not cross with other levels in the region which we consider, 
and the interaction between the lowest level and the others 
is small. This fact validates the adiabatic approximation to a large 
extent. 
Notice that 
each single-particle level is smoothly connected 
to those of two isolated fission fragments at large $Q_2$. 
In particular, the lowest level is connected to 
the lowest level of the heavier fragment
because the $\Lambda$ binding energy is larger in the 
heavier fragment. 
As a consequence, 
the $\Lambda$ particle is always stuck to the 
heavier fission fragment when it is at the lowest 
single-particle level.

\subsection{$\Lambda$ at a higher single-particle level}

Let us next discuss 
the fission process when 
a $\Lambda$ particle is at a higher single-particle level.
This corresponds to the case where 
the hypernucleus goes to fission 
before the $\Lambda$ particle produced experimentally, 
which is at first at an excited state, deexcites to the lowest level. 
For this purpose, 
we select four single-particle levels 
with $K=1/2$ around the neutron Fermi energy, $K$ being the projection of 
the single-particle angular momentum onto the symmetry axis. 
Those four levels are 
denoted as level A, B, C, and D 
in Fig.  6 by the thick solid lines. 
As in the previous subsection, we assume that 
$\Lambda$ particle adiabatically moves at the level crossing 
points. 

\begin{figure}
\begin{center}
\begin{tabular}{cc}
\includegraphics[width=0.46\textwidth,clip]{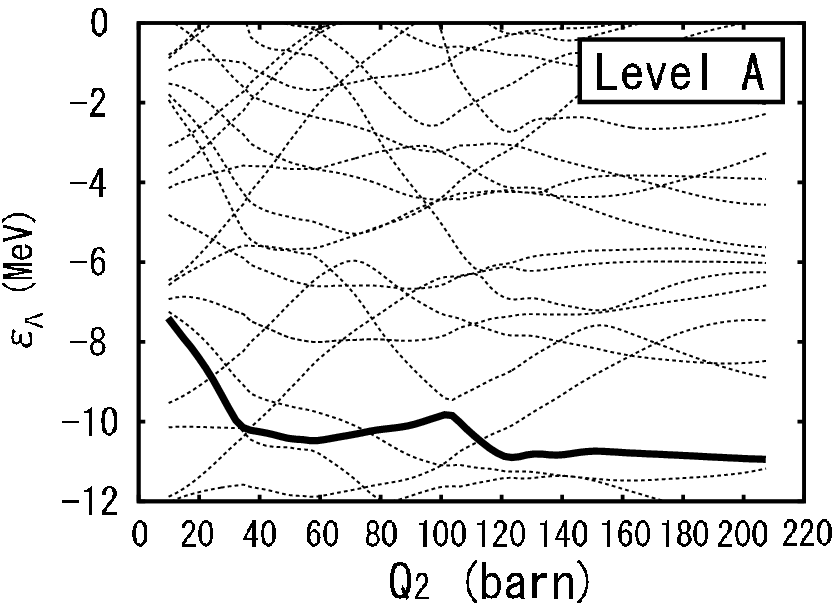}&
\includegraphics[width=0.46\textwidth,clip]{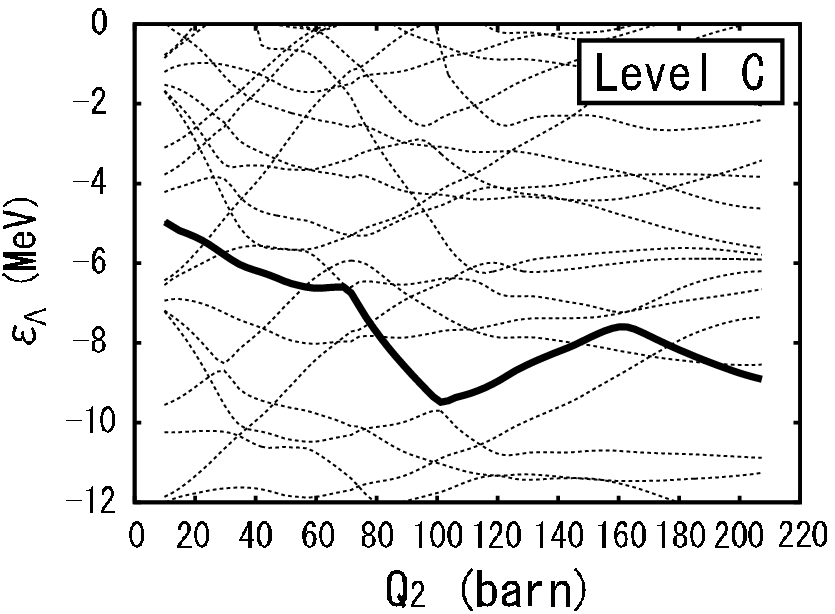}\\
\includegraphics[width=0.46\textwidth,clip]{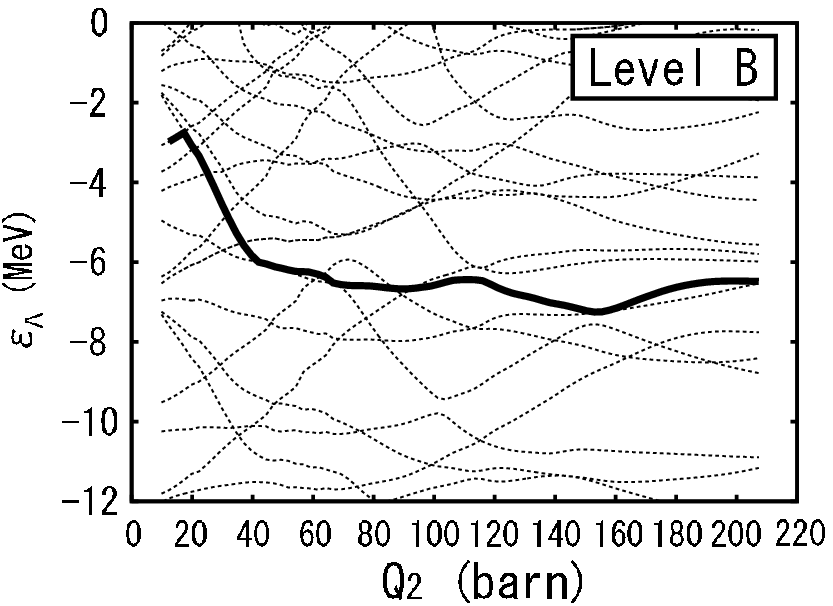}&
\includegraphics[width=0.46\textwidth,clip]{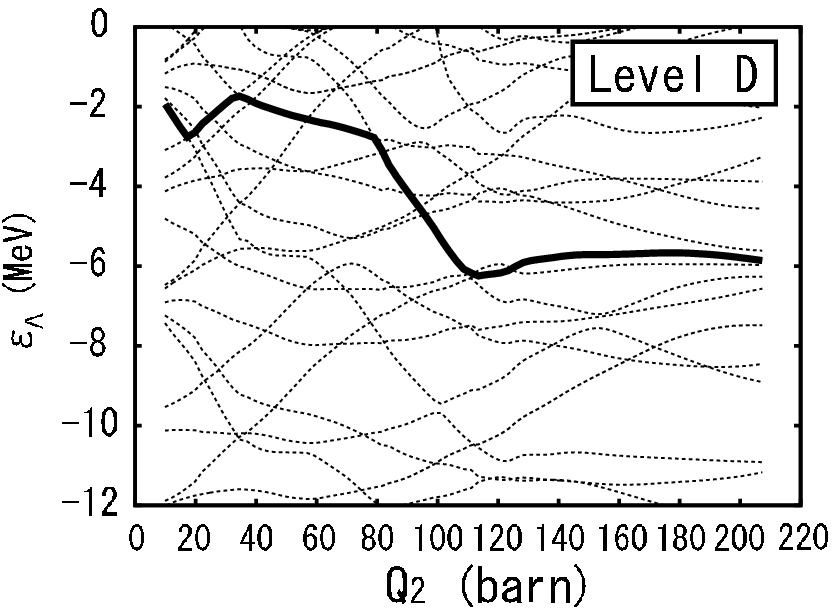}
\end{tabular}
\caption{
Same as Fig. 5, but for the cases when the 
$\Lambda$ particle is at four different $K$=1/2 levels 
denoted by the thick solid lines. 
}
\label{sp2}
\end{center}
\end{figure}

\begin{figure}
\begin{center}
\includegraphics[width=0.49\textwidth,clip]{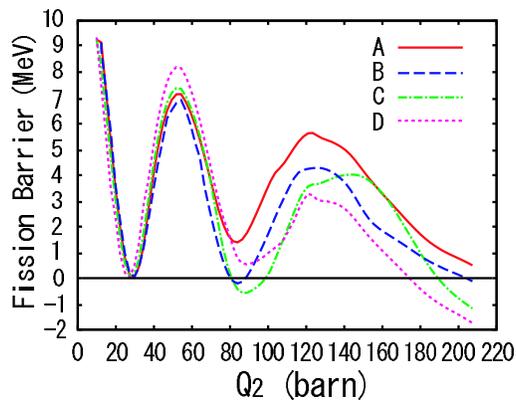}
\caption{Fission barrier for the $^{239}_{~~\Lambda}$U nucleus 
as a function of the total quadrupole moment $Q_2$ when 
the 
$\Lambda$ particle occupies the 
four different single-particle levels denoted by the thick solid line in 
Fig. 5. }
\label{bar2}
\end{center}
\end{figure}

The resultant fission barrier curves 
for each configuration are shown in Fig. \ref{bar2}.
The height for the inner and outer barriers is listed in Tab. \ref{Bf2}.
The fission barrier heights for all the cases are 
lower than that for $^{238}$U shown 
in Fig. \ref{bar1} (see also Table 1).  
As we argued in the previous subsection,
these changes of the barrier height with respect to 
the barrier height for the core nuclues ${}^{238}$U 
are intimately connected to 
the behaviour of $\Lambda$ single-particle energies.
In Fig. \ref{sp2}, 
one notices that 
the single-particle energies for the occupied levels 
decrease away from the ground state 
($Q_2=27.15$ b), 
leading to the decrease of fission barrier height.

\begin{table}
\begin{center}
\begin{tabular}{c|cccc}
\hline
                 & (A) & (B) & (C) & (D)\\
\hline
$B_f$(inner) (MeV) & $7.12$ & $6.87$ & $7.31$ & $8.16$ \\
$B_f$(outer) (MeV) & $5.61$ & $4.25$ & $3.95$ & $3.23$ \\
\hline
\end{tabular}
\label{Bf}
\caption{
The height of the inner and outer fission barriers for the 
$^{239}_{\;\;\;\Lambda}$U nucleus for the $\Lambda$ particle configurations 
shown in Fig. 6. 
}
\label{Bf2}
\end{center}
\end{table}
\begin{figure}
\begin{center}
\includegraphics[height=0.70\textwidth,clip]{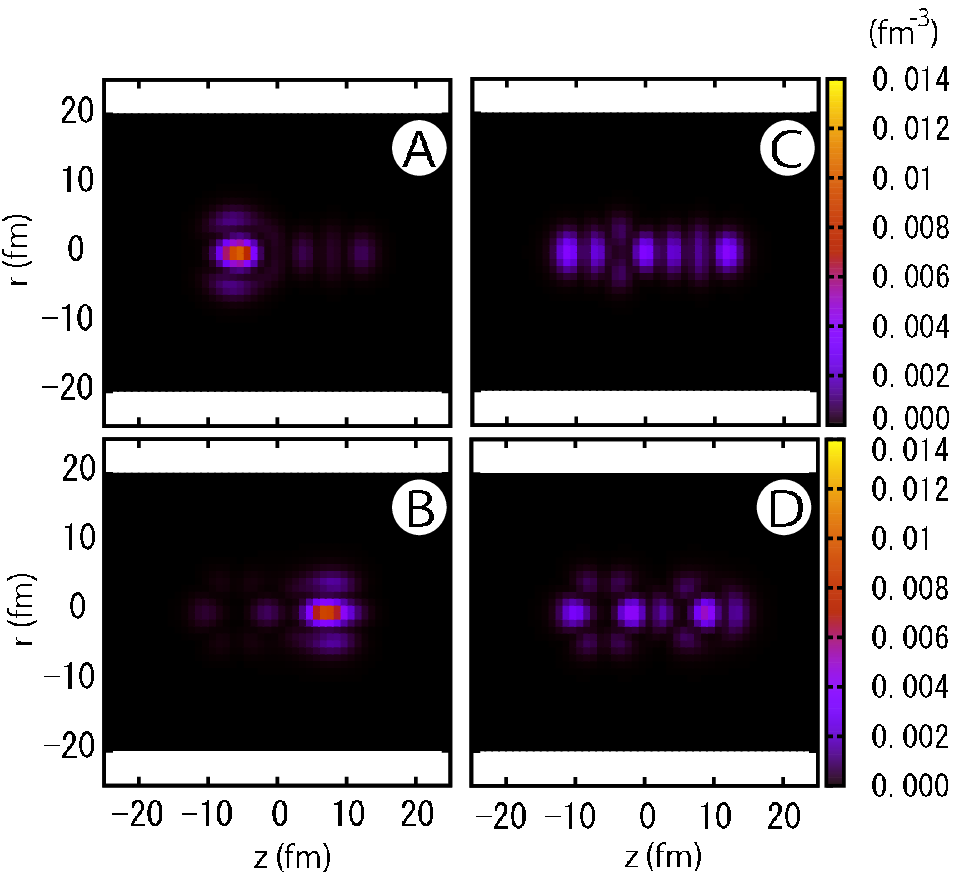}
\caption{
The density distribution for the $\Lambda$ particle at $Q_2=200$ b 
for the configurations $A$, $B$, $C$, and $D$.}
\label{density2}
\end{center}
\end{figure}

In order to see to which fragment the $\Lambda$ particle is stuck 
when it is at an excited level,
we plot the $\Lambda$ densities for the levels A, B, C, and D 
at $Q_2=200$ b in Fig. \ref{density2}.
We have confirmed that the density distribution for the 
core nucleus is almost the same as 
that with $\Lambda$ at the lowest level shown in Fig. 3.
In Fig. \ref{density2},
one sees that the $\Lambda$ particle is attracted by 
the heavier fragment ($z<0$) for the level A, 
similar to 
the case of $\Lambda$ particle at the lowest single-particle level.
On the other hand,
the $\Lambda$ particle moves 
to the lighter fragment ($z>0$) in the case of level B.
For the levels C and D, 
the density distribution is not localized and 
we cannot say to which fragment 
the $\Lambda$ particle moves. 

\begin{figure}
\begin{center}
\begin{tabular}{ccc}
&
\includegraphics[width=0.33\textwidth,clip]{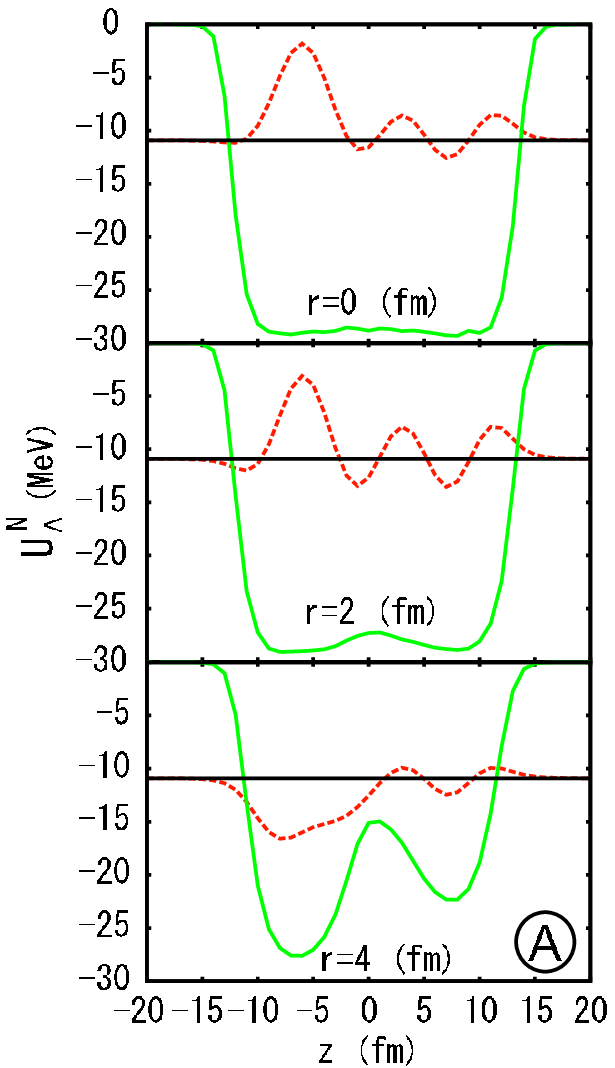}&
\includegraphics[width=0.33\textwidth,clip]{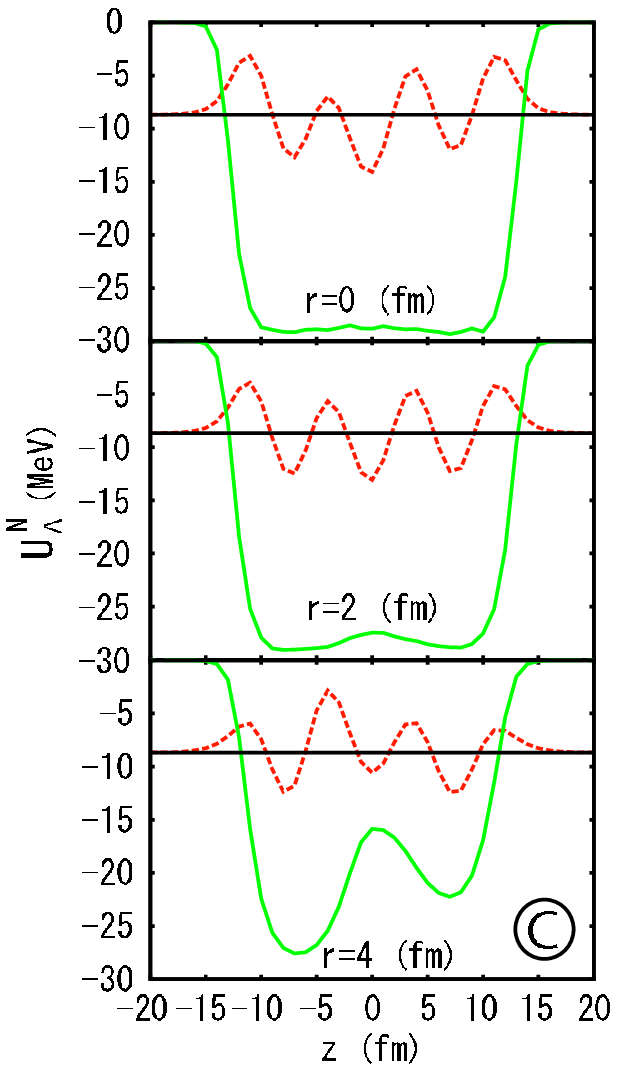}\\
\includegraphics[width=0.33\textwidth,clip]{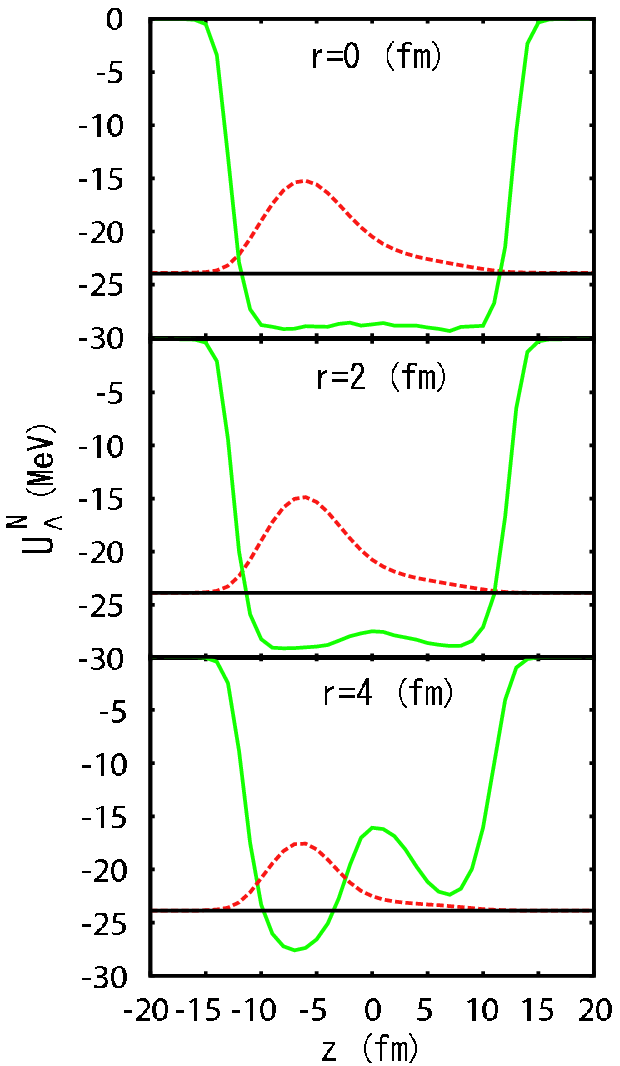}&
\includegraphics[width=0.33\textwidth,clip]{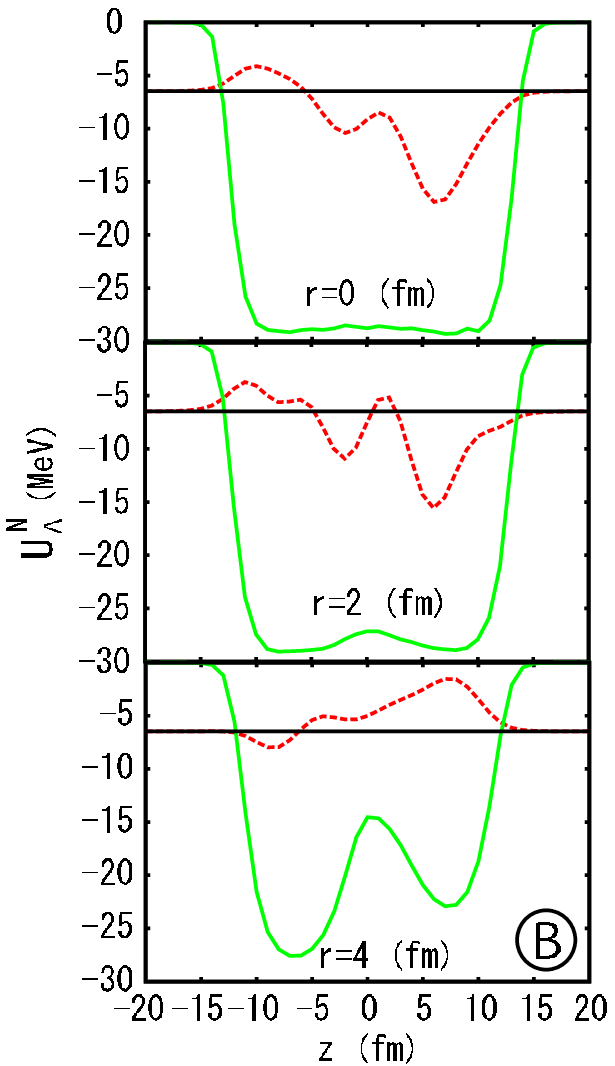}&
\includegraphics[width=0.33\textwidth,clip]{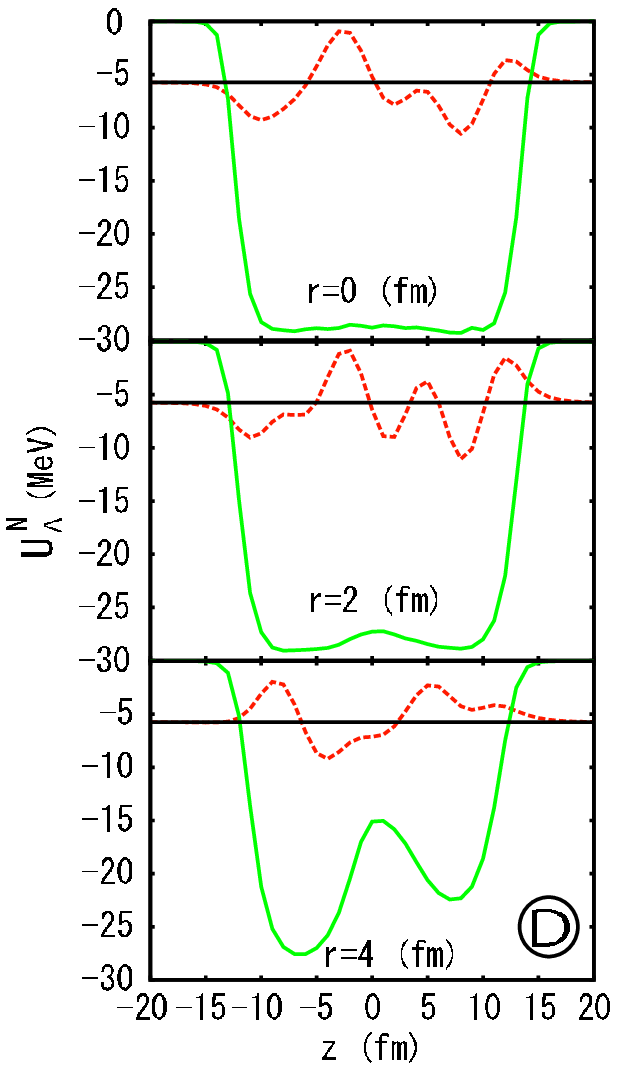}
\end{tabular}
\caption{
The mean-field potential for the $\Lambda$ particle, 
$U_\Lambda^N(\vec{r})$, (the solid line) and 
the $\Lambda$ wave function, $\phi_\Lambda(r,z)$, (the dashed line) 
at $r$=0, 2, and 4 fm and at the total quadrupole moment of $Q_2=200$ barn
for the lowest(the left bottom panel), A, B, C, and D levels.
}
\label{potwav}
\end{center}
\end{figure}

Fig. \ref{potwav} shows 
the $\Lambda$ mean-field potential $U_\Lambda^N(\vec{r})$ (the solid line) 
and the $\Lambda$ wave functions $\phi_\Lambda(r,z)$ (the dashed line) 
at $Q_2=200$ barn for the lowest, A, B, C, and D levels.
These are plotted for three different values of $r$, that is, $r$=0 fm, 2 fm, 
and 4 fm. The localization of the $\Lambda$ wave function for the lowest level, 
the levels A and B are clearly seen. 
For the levels C and D, the wave function is expected to be localized 
more clearly at larger 
values of quadrupole moment $Q_2$, after a few more level crossings, 
although we are unable to check it numerically because of 
a difficulty of numerical calculation at large deformations. 

\section{Conclusion}

We have calculated the fission barrier curve for 
${}^{239}_{\;\;\;\Lambda}$U hypernucleus 
with the constraint Skyrme-Hartree-Fock + BCS method
and compared it with that for $^{238}$U.
For this purpose, we have assumed that 
the $\Lambda$ particle adiabatically follows the fission process. 
We found that 
the fission barrier height increases when a $\Lambda$ particle is added
to the lowest single-particle level of $^{239}_{~~\Lambda}$U. 
We argued that this is caused because the 
the lowest $\Lambda$ single-particle energy 
increases towards the saddle point. 
On the other hand, we have confirmed that, when the single-particle energy 
decreases at the saddle point as compared to the ground state, the fission 
barrier height with that configuration decreases. 
We also discussed the effect of $\Lambda$ particle on 
the mass partition of fission fragments. 
We have shown that the mass of the fragment to which the $\Lambda$ 
particle is attached increases 
due to the attractive interaction between 
$\Lambda$ and nucleons. 

The adiabatic approximation which we employed in this paper would be 
justified when the $\Lambda$ particle occupies the lowest single-particle 
level. 
For a $\Lambda$ particle at an excited state, on the other hand, 
its validity depends on 
how fast the fission takes place. 
When the fission takes place rapidly, 
the $\Lambda$ particle may diabatically follow the fission process. 
In order to take into account the deviation from the adiabatic 
approximation, a dynamical calculation for fission with the Landau-Zener 
transition at level crossings is called for. 

Lastly, we would like to point some possible applications of the fission of 
heavy $\Lambda$ hypernuclei. One is a possibility to produce heavy 
neutron-rich $\Lambda$ hypernuclei. A light neutron-rich hypernucleus 
$^{10}_\Lambda$Li has been successfully produced via ($\pi^-,K^+$) reaction 
\cite{Saha05,Tamura09}. This reaction converts two protons to a 
$\Lambda$ and a neutron. It is therefore difficult with this method to 
produce a hypernucleus far from the stability line in the heavy mass region. 
In contrast, since the $\Lambda$ particle eventually emerges in one of the fission 
fragments, the fission of heavy $\Lambda$ hypernuclei may open up a novel 
method to produce a heavy neutron-rich $\Lambda$ hypernucleus. Notice that heavy 
neutron-rich nuclei (without hyperon) has been produced via in-flight 
fission of $^{238}$U at new generation radioactive isotope beam 
facilities\cite{RIBF08,MSU09}. 
Another 
application of fission of heavy $\Lambda$ hypernuclei is 
the nuclear transmutation. 
If a $\Lambda$ particle remains at a fission fragment, it eventually 
decays by weak interaction (predominantly by non-mesonic decay). 
The weak decay of $\Lambda$ induces fission, and/or neutron and proton emissions of the fission fragment 
which the $\Lambda$ particle originally sticks to\cite{Ohm97,Kulessa98}. 
It is an interesting future question whether 
such processes can be utilized to transmute 
long-lived radioactive wasts produced at nuclear power plants. 

\section*{Acknowledgement}
We thank Myaing Thi Win, E. Hiyama, 
and S. Hirenzaki for useful discussions.
This work was supported by the Japanese
Ministry of Education, Culture, Sports, Science and Technology
by Grant-in-Aid for Scientific Research under
the program number 19740115.

\end{document}